\documentclass[aps,prl,reprint,superscriptaddress,longbibliography]{revtex4-2}

\usepackage{personal}

\begin{document}

\title{Thermal Pure States for Systems with Antiunitary Symmetries\\ and Their Tensor Network Representations}

\author{Yasushi Yoneta}
\email{yasushi.yoneta@riken.jp}
\affiliation{Center for Quantum Computing, RIKEN, 2-1 Hirosawa, Wako, Saitama 351-0198, Japan}

\date{\today}

\begin{abstract}
Thermal pure state algorithms, which employ pure quantum states representing thermal equilibrium states instead of statistical ensembles, are useful both for numerical simulations and for theoretical analysis of thermal states. However, their inherently large entanglement makes it difficult to represent efficiently and limits their use in analyzing large systems. Here, we propose a new tensor network algorithm for constructing thermal pure states for systems with certain antiunitary symmetries, such as time-reversal or complex conjugate symmetry. Our method utilizes thermal pure states that, while exhibiting volume-law entanglement, can be mapped to tensor network states through simple transformations. Furthermore, our approach does not rely on random sampling and thus avoids statistical uncertainty. Moreover, we can compute not only thermal expectation values of local observables but also thermodynamic quantities. We demonstrate the validity and utility of our method by applying it to the one-dimensional XY model and the two-dimensional Ising model on a triangular lattice. Our results suggest a new class of variational wave functions for volume-law states that are not limited to thermal equilibrium states.
\end{abstract}

\maketitle

\textit{Introduction---}
Formulating statistical mechanics using pure states has attracted significant attention across various fields,
from quantum statistical physics~\cite{Popescu2006,Goldstein2006,Reimann2007,Sugiura2012,Sugiura2013,Hyuga2014}
to quantum gravity~\cite{Goto2021,Okuyama2021,Okuyama2024,Wei2024}.
This approach employs thermal pure states, which are locally indistinguishable from thermal equilibrium states, instead of statistical ensembles, to derive the thermal properties.
Thermal pure states are utilized not only for finite-temperature simulations~\cite{Yamaji2016,Shimokawa2016,Nasu2017,Endo2018,Tomishige2018,Suzuki2018,Oitmaa2018,Wietek2019,Hickey2019,Hickey2020,Suzuki2021,Shackleton2021,Powers2023,Coopmans2023,Davoudi2023} but also for theoretical analysis
to gain deep insights into the structure of thermal equilibrium states~\cite{Nakagawa2018,Fujita2018,Lu2019}.

The seminal works by Sugiura and Shimizu~\cite{Sugiura2012,Sugiura2013} provided a method for constructing thermal pure states, called the TPQ states, using Haar random states. They proved that the TPQ states yield thermal expectation values and thermodynamic quantities with an exponentially small probability of error in the thermodynamic limit. However, the complexity of Haar random states makes it difficult to represent efficiently, and numerical simulations are limited to systems of $30$--$40$ sites. Furthermore, statistical-mechanical quantities computed from the TPQ states have uncertainties due to their randomness, which become more pronounced at lower temperatures where the number of states decreases.

Thermal pure states must have a large amount of entanglement obeying a volume law because they are locally indistinguishable from the Gibbs state, whose von Neumann entropy of the reduced density matrix is equal to the thermodynamic entropy. This presents a major challenge in efficiently representing thermal pure states using variational wave functions. This is because tensor network states, such as matrix product states (MPS)~\cite{Fannes1989,Fannes1991,Verstraete2006,Verstraete2008,Schollwock2011}, which are the most widely used variational wave functions for quantum many-body states, require a small amount of entanglement to be represented with a small number of parameters.
Thus, algorithms such as the METTS method~\cite{White2009,Stoudenmire2010} and the TPQ-MPS method~\cite{Garnerone2013,Iwaki2021,Iwaki2022}, which cleverly decompose the statistical ensemble into a convex mixture of low-entanglement pure states, are also employed in practical applications.
However, to accurately recover the properties of the original ensemble, one needs to repeat the sampling of these states to reduce statistical uncertainties~\cite{Iwaki2024}.

In developing efficient algorithms, it is crucial to leverage symmetries. While unitary symmetries are often the primary focus, many realistic quantum many-body systems also possess antiunitary symmetries, such as time-reversal symmetry. It is known that antiunitary symmetries play a significant role in the study of quantum chaos~\cite{Mehta2004}. Given their importance, it is an intriguing question whether finite-temperature simulations can be fast-forwarded under such symmetries.

In a previous paper~\cite{Chiba2024_EAP}, the authors proposed a new type of thermal pure states, constructed in a manner quite different from existing thermal pure states. These states are obtained from well-structured states, termed entangled antipodal pair (EAP) states, rather than Haar random states. However, the structured nature of the EAP states was not fully exploited, restricting their numerical application to small systems. Furthermore, their applicability was limited to one-dimensional systems with complex conjugate symmetry.

The main contribution of this paper is to develop an efficient tensor network algorithm for constructing thermal pure states derived from EAP states, addressing the limitations of existing finite-temperature pure state algorithms in practical applications. To this end, we focus on the well-structured and deterministic natures of the EAP state, which facilitates the efficient construction and representation of the derived states and frees them from statistical uncertainty.
First, we generalize the EAP states to make our algorithm applicable to systems other than one-dimensional systems with complex conjugate symmetry. Next, we present thermal pure states constructed from these EAP states and provide formulas for statistical-mechanical quantities using these states. Furthermore, we develop a tensor network algorithm for efficiently constructing these states. Finally, we demonstrate the validity and usefulness of our method by applying it to paradigmatic models in one and two dimensions.

\begin{figure}
    \centering
    \begin{tabular}{lcr}
        \adjustbox{valign=b}{
            \subfloat[]{\includegraphics[keepaspectratio,width=.35\linewidth]{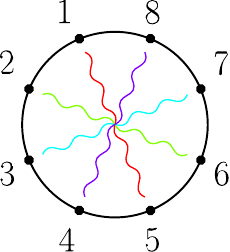}\label{fig:EAP_ring}}
        }
        & \ \ \  &
        \adjustbox{valign=b}{\begin{tabular}{@{}c@{}}
            \subfloat[]{\includegraphics[width=.55\linewidth]{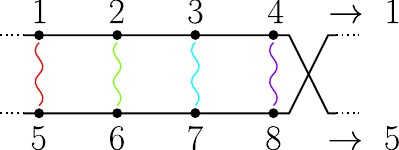}\label{fig:EAP_ladder}}\\
            \subfloat[]{\includegraphics[width=.55\linewidth]{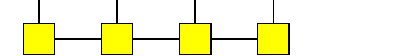}\label{fig:EAP_MPS}}
        \end{tabular}}
    \end{tabular}
    \caption{\protect\subref{fig:EAP_ring} Schematic diagram depicting an entangled antipodal pair state $\ket{\mathrm{EAP}(\hat{u})}$ for a one-dimensional system with periodic boundary conditions. Wavy lines represent the maximally entangled states $\ket{\Phi(\hat{u})}$.
    \protect\subref{fig:EAP_ladder}
    A system on a ring can be mapped to one on a ladder with M\"obius boundary conditions while preserving the short-range nature of the interactions. In this correspondence, the EAP state is mapped to a state with only local entanglement.
    \protect\subref{fig:EAP_MPS} Therefore, it can be represented as an MPS with a local dimension $4$ and a bond dimension $1$.}
    \label{fig:EAP}
\end{figure}

\textit{EAP state---}
We consider a closed quantum system composed of spins on a $d$-dimensional lattice $\Lambda = \prod_{i=1}^d \{1,2,\cdots,L_i\}$ with periodic boundary conditions. We write the system size as $N = \prod_{i=1}^d L_i$. We assume that the side length $L_1$ in the first direction is even. Let $d(r, r')$ denote the distance between two points $r, r' \in \Lambda$ on the lattice. Let $\hat{\sigma}^\mu (\mu=x,y,z)$ be the Pauli matrices, and $\ket{0}$ and $\ket{1}$ be the eigenvectors of $\hat{\sigma}^z$. We call the basis of the whole Hilbert space formed by the tensor products of $\ket{0}$ and $\ket{1}$ the spin basis.

As for the preparation, we define the antipodal site for each lattice site. Let the mapping $p$ that assigns each lattice site $r\in\Lambda$ to its antipodal site $p(r) (\in \Lambda)$.
In this paper, we select the simplest choice as~\footnote{For our results, it is sufficient that $p$ satisfies the following conditions: (1) $p$ is the spatial translation on $\Lambda$, (2) $p(p(r)) = r$, and (3) the minimum distance between antipodal pairs is $\Theta(N^{1/d})$.
It is worth noting that whereas $p$ is unique in one-dimensional systems, it can be chosen in various ways in higher-dimensional systems.}
\begin{align}
    p(r_1, r_2, \cdots, r_d)
    = (r_1 + L_1/2 \mod L_1, r_2, \cdots, r_d).
\end{align}
With this choice, $p(r)$ corresponds to the antipodal point of $r$ on the ``meridian'' of the $d$-dimensional torus.
We can divide the lattice $\Lambda$ into the ``left half'', denoted as $\Lambda_L$, and its complement, ensuring that each subsystem contains exactly one site from each antipodal pair. Specifically, $\Lambda_L$ is defined as
\begin{align}
    \Lambda_L = \left\{ (r_1,r_2,\cdots,r_d) \in \Lambda \middle| 1 \leq r_1 \leq L_1/2 \right\}.
\end{align}

Now, we define the EAP state for the spin-$1/2$ system on $\Lambda$~\footnote{Here, we consider only EAP states that are invariant (up to a phase factor) under spatial translations, as required for subsequent applications.}. Let $\hat{u}$ be a $2 \times 2$ unitary matrix that is either symmetric or antisymmetric,
\begin{align}
    \hat{u}^\mathrm{T} = (-1)^k \hat{u} \quad (k \in \mathbb{Z}), \label{eq:u-transpose}
\end{align}
where the superscript $\mathrm{T}$ denotes the transpose in the $\{\ket{0},\ket{1}\}$ basis.
As explained below Eq.~\eqref{eq:antiunitary-symmetry},
$\hat{u}$ is chosen according to the antiunitary symmetry of the system. Then we define the EAP state associated with $\hat{u}$ as~\footnote{Interestingly, some EAP states coincide with lattice theory counterparts of crosscap states in $(1+1)$-dimensional CFTs, studied in the contexts of quantum field theory and mathematical physics~\cite{Ishibashi1989,Caetano2022,Ekman2022,Guica2015,Maloney2016,Wei2024_XCFT}. However, we consider more general systems beyond $(1+1)$-dimensional CFTs. Moreover, even when crosscap states can be defined, EAP states generally do not coincide with them. Therefore, to avoid confusion, we refer to $\ket{\mathrm{EAP}(\hat{u})}$ as the EAP state.}
\begin{align}
    \ket{\mathrm{EAP}(\hat{u})}
    = \bigotimes_{r \in \Lambda_L} \ket{\Phi(\hat{u})}_{r,p(r)}.
    \label{eq:def_EAP}
\end{align}
Here $\ket{\Phi(\hat{u})}_{r,p(r)}$ is a maximally entangled state between antipodal pairs of spins at sites $r$ and $p(r)$ defined by~\footnote{If we choose $\hat{u}$ to be the identity operator $\hat{I}$ or one of the Pauli matrices $\hat{\sigma}^\mu (\mu=x,y,z)$, then in each case, $\ket{\Phi(\hat{u})}_{r,p(r)}$ reduces to one of the four Bell states. Thus, Eq.~\eqref{eq:def_EAP} is an extension of the definition of the EAP state presented in Ref.~\cite{Chiba2024_EAP}.}
\begin{align}
    \ket{\Phi(\hat{u})}_{r,p(r)}
    = \hat{I} \otimes \hat{u}
    \left[
        \ket{0}_{r}\ket{0}_{p(r)}+\ket{1}_{r}\ket{1}_{p(r)}    
    \right],
\end{align}
where $\hat{I}$ and $\hat{u}$ act on site $r$ and $p(r)$, respectively.
Although this state might appear to be macroscopically inhomogeneous, it is invariant under translations (up to the sign) because $\ket{\Phi(\hat{u})}_{r,p(r)}$ is (anti)symmetric under the exchange of $r$ and $p(r)$, i.e., $\mathrm{Swap}_{r,p(r)}\ket{\Phi(\hat{u})}_{r,p(r)}=(-1)^k\ket{\Phi(\hat{u})}_{r,p(r)}$.

One remarkable property of the EAP state is that it is locally indistinguishable from the Gibbs state
\begin{align}
    \hat{\rho}_{N}^\mathrm{can}(\beta)
    = \frac{e^{-\beta\hat{H}}}{Z}
\end{align}
at the inverse temperature $\beta=0$~\cite{Chiba2024_EAP}. Here $Z = \mathrm{Tr} [ e^{-\beta\hat{H}} ]$ is a partition function. In fact, the reduced density matrix of the EAP state for any subsystem $X(\subset\Lambda)$ with diameter $D(X)=\max_{r,r' \in X} d(r,r')$ less than $L_1/2$ coincides with the maximally mixed state $\propto \hat{I}_X$, which equals that for the Gibbs state at $\beta=0$. Thus, EAP states are thermal pure states at infinite temperature.
In the following, we present a method to construct thermal pure states at finite temperatures from EAP states.

\begin{figure*}[!t]
    \centering
    \subfloat[correlation function $\braket{\hat{\sigma}_{i}^x\hat{\sigma}_{i+j}^x}$]{\includegraphics[height=5.0cm]{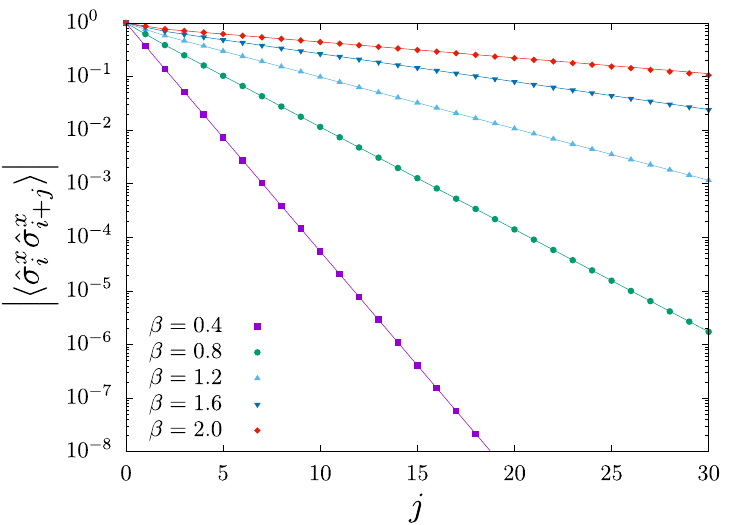}\label{fig:XY_correlation}} \hfill
    \subfloat[free energy density $f$]{\includegraphics[height=5.0cm]{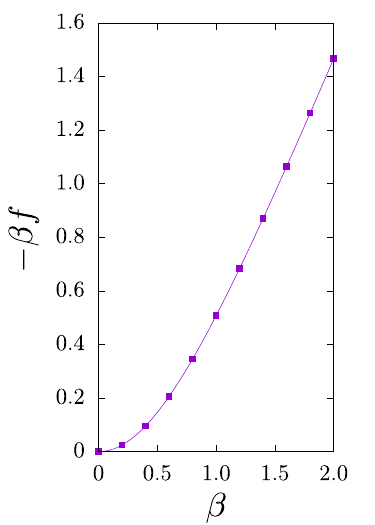}\label{fig:XY_logZ}} \hfill
    \subfloat[entanglement entropy $S_\mathrm{vN}$]{\includegraphics[height=5.0cm]{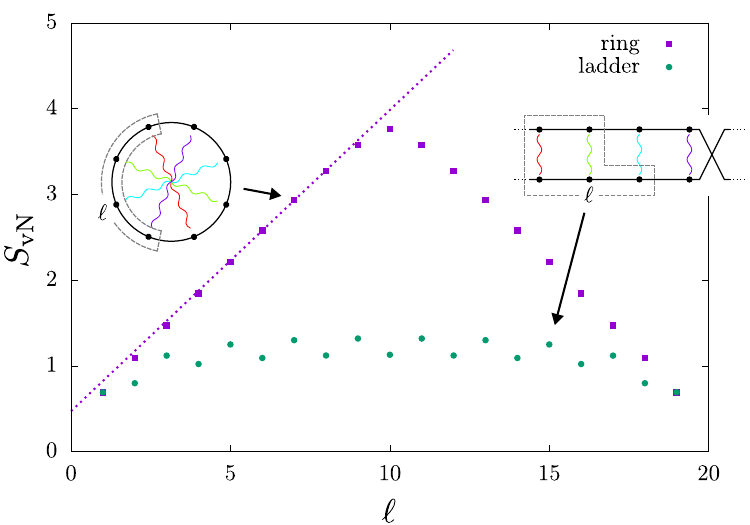}\label{fig:XY_EE}}
    \caption[]{%
        Properties of the {\BBB} state for the one-dimensional XY model with $J^{xx}=1, J^{yy}=1/2$.
        \subref{fig:XY_correlation} Two-point correlation function $\braket{\hat{\sigma}_{i}^x\hat{\sigma}_{i+j}^x}$ and \subref{fig:XY_logZ} free energy density $f$ for a $200$-site chain. The solid lines represent the exact values in the thermodynamic limit.
        \subref{fig:XY_EE} Entanglement entropy for a $20$-site chain at the inverse temperature $\beta=1$ as a function of subsystem size $\ell$. The system is partitioned in two different ways: (purple squares) by cutting the original ring and (green circles) by cutting the ladder obtained by rearrangement.
        The dotted line represents the volume law whose coefficients corresponds to the exact value of the thermodynamic entropy density at the same temperature.
        Imaginary-time evolution is carried out using a second-order Trotter decomposition with an imaginary time step of $0.05$ and a truncation threshold of $10^{-14}$.
    }
    \label{fig:XY}
\end{figure*}

\textit{Imaginary-time evolved EAP state---}
We can generate a finite-temperature thermal pure state from an EAP state. In doing so, the EAP state must be chosen appropriately according to the symmetry of the system.

Consider a system with translation-invariant short-range interactions. We assume that the Hamiltonian $\hat{H}$ is invariant
\begin{align}
    \hat{\Theta}^\dagger \hat{H} \hat{\Theta} = \hat{H}
    \label{eq:antiunitary-symmetry}
\end{align}
under an antiunitary transformation expressed as $\hat{\Theta} = \hat{u}^{\otimes N} \hat{K}$, where $\hat{u}$ is a $2 \times 2$ unitary matrix satisfying ${\hat{u}}^\mathrm{T}=(-1)^k\hat{u}$, and $\hat{K}$ represents the complex conjugation with respect to the spin basis.
Hamiltonians satisfying Eq.~\eqref{eq:antiunitary-symmetry} can be classified into following two classes:
\begin{enumerate}
    \item Hamiltonians whose matrix representations with respect to the spin basis are real, $\hat{H} = \hat{H}_\mathrm{real}$ (the case of $\hat{u} = \hat{I}$), and their conjugations by global $\mathrm{SU}(2)$ spin rotations $\hat{W}=\hat{w}^{\otimes N}$, $\hat{H} = \hat{W}^\dagger \hat{H}_\mathrm{real} \hat{W}$ (the case of $\hat{u}=\hat{w}\hat{w}^\mathrm{T}$),
    \item Hamiltonians with time-reversal symmetry (the case of $\hat{u} = i \hat{\sigma}^y$).
\end{enumerate}
These classes include various models of interest in statistical mechanics.

We define the {\bbb} ({\BBB}) state at the inverse temperature $\beta$ as an unnormalized state obtained by evolving the EAP state associated with $\hat{u}$ in imaginary time $\beta/4$:
\begin{align}
    \ket{\beta}
    = e^{-\frac{1}{4}\beta\hat{H}} \ket{\mathrm{EAP}(\hat{u})}.
\end{align}
Then, for any local observable $\hat{O}$, this state is indistinguishable from the Gibbs state $\hat{\rho}_{N}^\mathrm{can}(\beta)$ at the same inverse temperature $\beta$~\footnote{See Supplemental Material for the derivation.}:
\begin{align}
    \lim_{N\to\infty} \mathrm{Tr} \left[ \hat{\rho}_{N}^\mathrm{can}(\beta) \hat{O} \right]
    = \lim_{N\to\infty} \frac{\braket{\beta|\hat{O}|\beta}}{\braket{\beta|\beta}}.
    \label{eq:formula_expectation-value}
\end{align}
In other words, $\ket{\beta}$ is a thermal pure state at the inverse temperature $\beta$.

Moreover, one can also calculate the free energy density $f$ from the norm of the {\BBB} state as~\cite{Note5}
\begin{align}
    - \beta f(\beta)
    = \lim_{N\to\infty} \frac{1}{N} \log Z
    = \lim_{N\to\infty} \frac{2}{N} \log \braket{\beta|\beta}.
    \label{eq:formula_free-energy}
\end{align}
Therefore, one can obtain all thermodynamic quantities, including the thermodynamic entropy and specific heat, as these quantities can be computed from the free energy~\cite{Callen1985}.

Here, we should mention the distinction between the {\BBB} state and a purified Gibbs state~\cite{Verstraete2004,Feiguin2005}.
The purified Gibbs state is a pure state of an extended system that includes an ancillary system, and when focusing only on the target system, it is completely identical to the Gibbs state in all physical properties. As the Gibbs state is characterized as the state that maximizes the von Neumann entropy~\cite{Jaynes1957_1,Jaynes1957_2}, it is a highly mixed state. Therefore, it cannot be used to analyze properties unique to pure states.
In contrast, the {\BBB} state is a pure state of a closed system and thus serves as a theoretical example of a pure state in thermal equilibrium, providing access to properties of such a state, including the entanglement structure~\cite{Nakagawa2018,Fujita2018}.

\textit{Tensor network method---}
As discussed in the introduction, thermal pure states generally exhibit volume-law entanglement and therefore cannot be efficiently represented as tensor network states. However, {\BBB} states, despite also obeying a volume law, can be brought into forms that can be efficiently represented as tensor network states through simple transformations due to the well-structured nature of EAP states. This makes them useful for numerical calculations.

Let us describe a tensor network method for constructing the {\BBB} state, taking an example of a one-dimensional system $\Lambda=\{1,2,\cdots,N\}$ (Fig.~\ref{fig:EAP_ring}).
Apparently, the EAP state contains a large amount of long-range entanglement between sites separated by $N/2$ (Fig.~\ref{fig:EAP_ring}). To address this, we map the ring to a ladder of length $N/2$
(Fig.~\ref{fig:EAP_ladder}):
\begin{align}
    &\Lambda \ni j
    \mapsto
    \begin{cases}
        [A,j] & (j \leq N/2)\\
        [B,j-N/2] & (\text{otherwise})
    \end{cases},
\end{align}
where $[A,j]$ and $[B,j]$ denote the $j$-th site on the upper and lower chains of the ladder, respectively~\cite{Li2020,Tang2020}. Here, to ensure that the system is mapped to a translation-invariant and short-range interacting system on the ladder, we adopt M\"obius boundary conditions:
\begin{align}
    [A,N/2+1] = [B,1], \quad
    [B,N/2+1] = [A,1].
\end{align}
Then the antipodal pair ($j$ and $p(j)=j+N/2$) on the ring maps to a pair of spins ($[A,j]$ and $[B,j]$) aligned vertically on the ladder, so the maximally entangled state $\ket{\Phi(\hat{u})}_{j,p(j)}$ is local on the ladder.
Thus, by treating the vertically aligned pair of spins
as a single site $\cong \mathbb{C}^4$, the EAP state can be regarded as a product state without any spatial entanglement and can be represented as an MPS with a bond dimension $1$ (Fig.~\ref{fig:EAP_MPS}).
Consequently, the {\BBB} state, which is the imaginary-time evolution of this state
with a Hamiltonian where the interactions are short ranged (except for boundary terms introduced to impose the M\"obius boundary conditions~\footnote{These boundary terms are necessary to describe a translation-invariant system with periodic boundary conditions on the original lattice. While they are unfavorable for numerical calculations using MPS, they offer the advantage of exponentially suppressing finite-size effects compared to cases where boundary terms are absent and open boundary conditions are imposed~\cite{Iyer2015,Chiba2024_EAP}.}),
is expected to contain only a small amount of entanglement and can be efficiently represented as an MPS, as demonstrated in the example below.
Thus, by performing the imaginary-time evolution with a small time step $\delta\beta/4$ using the time-evolving block decimation algorithm~\cite{Vidal2004,Verstraete2004,White2004,Daley2004}, one can efficiently obtain a series of {\BBB} states at the inverse temperatures $0,\delta\beta,2\delta\beta,\cdots$.

\textit{Application~1: one-dimensional XY model---}
To demonstrate the validity and utility of the {\BBB} state, we apply our method to the one-dimensional XY model, whose exact solutions are known~\cite{Lieb1961}, defined by the Hamiltonian
\begin{align}
    \hat{H}
    = \sum_{j=1}^N \left[
        J^{xx} \hat{\sigma}_{j}^x \hat{\sigma}_{j+1}^x
      + J^{yy} \hat{\sigma}_{j}^y \hat{\sigma}_{j+1}^y
    \right].
    \label{eq:XY}
\end{align}
Since this Hamiltonian has time-reversal symmetry, we can choose $\hat{u}=i\hat{\sigma}^y$ to satisfy Eq.~\eqref{eq:antiunitary-symmetry} and generate the {\BBB} state from the EAP associated with this $\hat{u}$~\footnote{Since it is also a real matrix, we can also choose $\hat{u}=\hat{I}$. Both choices yield the same results in the thermodynamic limit.}.
Following the algorithm explained above, we successfully constructed the {\BBB} state for $N=200$, a significant improvement compared to the TPQ algorithm, which is limited to about $30$ sites.

Figure~\ref{fig:XY_correlation} shows the two-point correlation function $\braket{\hat{\sigma}_{i}^x\hat{\sigma}_{i+j}^x}$ calculated using the {\BBB} state. The results are in good agreement with the exact solution in the thermodynamic limit, confirming that the {\BBB} state is a thermal pure state (and, of course, free from statistical uncertainty). Furthermore, as shown in Figure~\ref{fig:XY_logZ}, the free energy calculated using the {\BBB} state also agrees with the exact value.

We also investigate the entanglement properties of the {\BBB} state.
Figure~\ref{fig:XY_EE} shows the entanglement profile of the {\BBB} state when the system is partitioned in two different ways.
When the system is divided into two subsystems by cutting the original ring, the entanglement entropy between them obeys a volume law.
Moreover, it can be seen that the coefficient of the volume law agrees with the thermodynamic entropy density. This not only corroborates the {\BBB} state as a thermal pure state but also enables us to extract the thermodynamic entropy from the entanglement scaling.
In contrast, when the system is divided by cutting the ladder after rearrangement, the entanglement entropy obeys an area law. This enables the efficient representation of the {\BBB} state as an MPS and dealing with large systems.

\begin{figure}
    \centering
    \subfloat[EAP state for a triangular lattice]{\includegraphics[width=\linewidth]{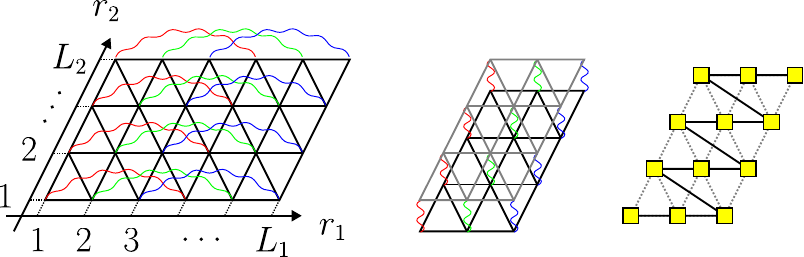}\label{fig:triangular-lattice}}\\
    \begin{tabular}{cc}
        \adjustbox{valign=b}{
            \subfloat[energy density $u$]{\includegraphics[keepaspectratio,height=.45\linewidth]{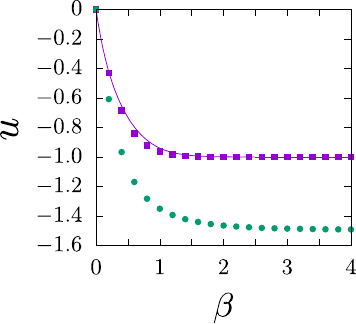}\label{fig:triangular-Ising_u}}
        }
        \adjustbox{valign=b}{
            \subfloat[free energy density $f$]{\includegraphics[keepaspectratio,height=.45\linewidth]{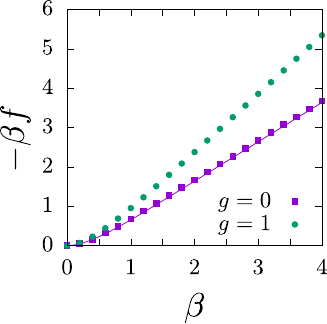}\label{fig:triangular-Ising_logZ}}
        }
    \end{tabular}
    \caption{%
        \protect\subref{fig:triangular-lattice} Schematic diagram depicting the triangular lattice and the EAP state employed in our numerical calculations. Wavy lines represent the maximally entangled state.
        \protect\subref{fig:triangular-Ising_u} Energy density $u=\braket{\hat{H}}/N$ and \protect\subref{fig:triangular-Ising_logZ} free energy density $f$ obtained from the {\BBB} state for the two-dimensional transverse-field Ising model on the triangular lattice with $J=1$, $L_1=6$, and $L_2=10$.
        Imaginary-time evolution is carried out using a second-order Trotter decomposition with an imaginary time step of $0.05$ and a truncation threshold of $10^{-8}$.
    }
    \label{fig:triangular-Ising}
\end{figure}

\textit{Application~2: two-dimensional Ising model on a triangular lattice---}
Our algorithm can be directly generalized to two-dimensional systems. Let us apply the {\BBB} state to the antiferromagnetic transverse-field Ising model on a triangular lattice of size $N = L_1 \times L_2$ with periodic boundary conditions (left of Fig.~\ref{fig:triangular-lattice}). The Hamiltonian is given by
\begin{align}
    \hat{H}
    = J \sum_{\langle r, r' \rangle} \hat{\sigma}_{r}^z \hat{\sigma}_{r'}^z
    - g \sum_{r} \hat{\sigma}_{r}^x
    \quad (J>0),
\end{align}
where $\langle r, r' \rangle$ denotes the nearest-neighbor pairs.
This model reduces to the classical Ising model as $g \to 0$, whose exact results at finite temperature have been derived for $N\to\infty$~\cite{Wannier1950}.
Since this Hamiltonian is real in the spin basis, choosing $\hat{u}=\hat{I}$ satisfies Eq.~\eqref{eq:antiunitary-symmetry}. Then, using this $\hat{u}$, we define the EAP state and construct the {\BBB} state. In numerical calculations, similar to the one-dimensional case, we fold the lattice so that the antipodal sites are adjacent to each other (center of Fig.~\ref{fig:triangular-lattice})
and represent the EAP and {\BBB} states as a snake MPS on the folded lattice (right of Fig.~\ref{fig:triangular-lattice}).

Figures~\ref{fig:triangular-Ising_u} and \ref{fig:triangular-Ising_logZ} show the $\beta$-dependence of the energy density and free energy calculated from the {\BBB} state, respectively. Firstly, in the case of $g=0$, it is confirmed that the results for the {\BBB} state agree with the exact values. Additionally, numerical calculations are also feasible for $g=1$, successfully reaching $60$ sites. 
If we were to keep the state vector in a brute-force way, similar to the TPQ algorithm, it would require an enormous amount of memory, on the order of $10^7 \mathrm{\ TB}$. However, due to the well-structured nature of the EAP state underlying the {\BBB} state, it is possible to investigate the thermal properties of large systems using only about $10^0 \mathrm{\ GB}$ of memory.

\textit{Conclusion---}
We have extended the method of constructing finite-temperature thermal pure states based on EAP states, originally provided only for one-dimensional systems with complex conjugate symmetry, to systems with time-reversal symmetry or complex conjugate symmetry on general lattices. Using these states, we can calculate not only thermal expectation values of local observables but also thermodynamic quantities. Furthermore, we have developed an efficient numerical method using tensor networks for these states. Our algorithm allows for a much more efficient representation on a classical computer compared to the TPQ algorithm. In addition, since our algorithm does not rely on random sampling, it does not suffer from statistical uncertainty and enables us to calculate all statistical-mechanical quantities accurately only from a single state. We have demonstrated the validity and utility of our method by applying it to the one-dimensional XY model and the two-dimensional Ising model on a triangular lattice.

There are several directions for future work.
First, it is an important task to extend the applicability of our algorithm, which is currently restricted to systems with certain types of antiunitary symmetries.
Second, developing an efficient algorithms for finite-temperature nonequilibrium simulations using the {\BBB} state is also an important challenge. While the {\BBB} state can correctly describe dynamics starting from an equilibrium state, as time evolves, it loses its simple entanglement structure, making it difficult for naive methods to access long timescales~\cite{Note5}.
Third, our results suggest the high expressivity of states obtained by applying matrix product operators to EAP states. This implies the potential to use these states as variational wave functions for a broader class of volume-law states beyond thermal pure states. Therefore, investigating the expressivity of states derived from EAP states is also an interesting challenge.
Finally, EAP states could be applied to quantum algorithms. This is because EAP states can be easily constructed using single-depth quantum circuits with local two-qubit gates by rearranging qubits similarly to the algorithm presented in this paper. This contrasts with the Haar random state, which is hard to construct~\cite{Nakata2017,Nakata2021,Poulin2011}.

\textit{Acknowledgments---}
We are grateful to H.-H.~Tu, A.~Shimizu, Y.~Chiba, and Z.~Wei for useful discussions.
The MPS calculations in this work were performed using the ITensor library~\cite{itensor}.
YY was supported by the Special Postdoctoral Researchers Program at RIKEN.

\bibliography{document}

\newcommand{\beginsupplement}{%
	\setcounter{table}{0}
	\renewcommand{\thetable}{S\arabic{table}}%
	\setcounter{figure}{0}
	\renewcommand{\thefigure}{S\arabic{figure}}%
	\setcounter{section}{0}
	\renewcommand{\thesection}{\Roman{section}}%
	\setcounter{equation}{0}
	\renewcommand{\theequation}{S\arabic{equation}}%
    \setcounter{page}{1}
}
\clearpage
\onecolumngrid
\beginsupplement

\begin{center}
	\textbf{\large Supplemental Material for\\ ``Thermal Pure States for Systems with Antiunitary Symmetries\\ and Their Tensor Network Representations''}
\end{center}
\vspace{2mm}

\section{Derivation of formulas for statistical-mechanical quantities}
In this supplemental material, we derive formulas for calculating statistical-mechanical quantities from {\BBB} states under plausible assumptions (shown in Eqs.~\eqref{eq:assumption_expectation-value} and \eqref{eq:assumption_free-energy} below).

\subsection{Truncated imaginary-time evolution}
We begin by decomposing the Hamiltonian into three parts: the part that acts only on the ``left half'' $\Lambda_L$, the part that acts only on the ``right half'' $\Lambda_R = \Lambda \setminus \Lambda_L$, and the interaction between these two subsystems.
This decomposition is not unique in general. Here, we consider the following decomposition by defining the interaction $\hat{H}_\mathrm{CT}$ as a counter term that ensures the equality:
\begin{align}
    \hat{H} = \hat{H}_{N/2} \otimes \hat{I}_R + \hat{I}_L \otimes \hat{H}_{N/2} + \hat{H}_\mathrm{CT}.
    \label{eq:Hamiltonian-decomposition}
\end{align}
Here, $\hat{H}_{N/2}$ is the Hamiltonian on a lattice $\Lambda_{N/2} = \{1,2,\cdots,L_1/2\} \times \{1,2,\cdots,L_2\} \times \{1,2,\cdots,L_3\} \times \cdots \times \{1,2,\cdots,L_d\}$, whose number of lattice sites is $N/2$, with periodic boundary conditions. That is, $\hat{H}_{N/2}$ is the Hamiltonian for the same system as for $\hat{H}$, except that the side length of the lattice in the first direction is half that of the original lattice.
To explain the properties of $\hat{H}_\mathrm{CT}$, we virtually deform the lattice by ``pinching'' the torus so that for all $r_i \in \{1,2,\cdots,L_i\} \ (i=2,3,\cdots,d)$, lattice points $(1,r_2,r_3,\cdots,r_d)$ and $(L_1/2+1,r_2,r_3,\cdots,r_d)$ are adjacent to $(L_1/2+1,r_2,r_3,\cdots,r_d)$ and $(L_1,r_2,r_3,\cdots,r_d)$, respectively:
\begin{figure}[H]
    \centering
    \includegraphics{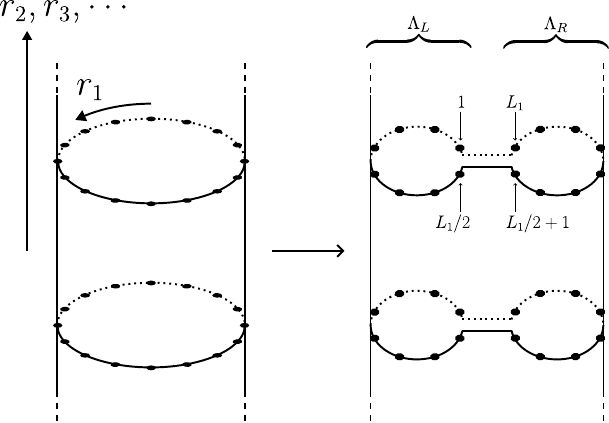}
\end{figure}
\noindent
For simplicity, we illustrate this by fixing the coordinates of the lattice other than $r_1$ ($r_2, r_3, \cdots$):
\begin{figure}[H]
    \centering
    \includegraphics{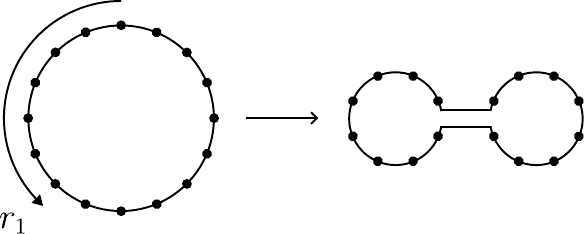}
\end{figure}
\noindent
This deformation localizes the boundary between subsystems $\Lambda_L$ and $\Lambda_R$. Consequently, for systems with short-ranged and translation-invariant interactions, the support of each term in Eq.~\eqref{eq:Hamiltonian-decomposition} is as follows:
\begin{figure}[H]
    \centering
    \includegraphics{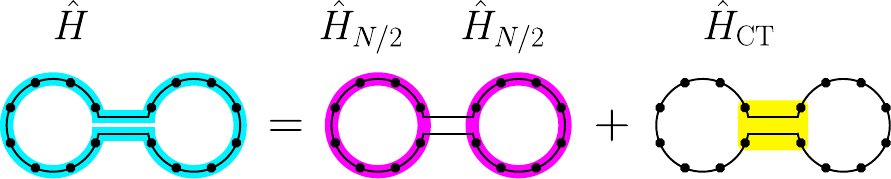}
\end{figure}
\noindent
Thus, $\hat{H}_\mathrm{CT}$ is an operator supported on a hyperplane of codimension $1$ in the deformed lattice (a point for one-dimensional cases and a line for two-dimensional cases), whose volume is negligible compared to the total volume in the thermodynamic limit.

In addition, on the deformed lattice, the entanglement in the EAP state $\ket{\mathrm{EAP}(\hat{u})}$ between a spin near the boundary and its antipodal pair is localized, and $\ket{\mathrm{EAP}(\hat{u})}$ is in a product state with elsewhere:
\begin{figure}[H]
    \centering
    \includegraphics{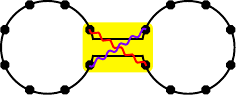}
\end{figure}
\noindent
Consequently, in the imaginary-time evolution of $\ket{\mathrm{EAP}(\hat{u})}$, it is anticipated that the presence of $\hat{H}_\mathrm{CT}$ will not significantly affect the regions far from the boundary between $\Lambda_L$ and $\Lambda_R$, as it does not propagate through correlations in the initial state.

Then we consider the imaginary-time evolution generated by the Hamiltonian without $\hat{H}_\mathrm{CT}$ and introduce an approximation $\ket{\tilde{\beta}}$ for $\ket{\beta}$ as
\begin{align}
    \ket{\tilde{\beta}} = e^{- \frac{1}{4} \beta [\hat{H}-\hat{H}_\mathrm{CT}]} \ket{\mathrm{EAP}(\hat{u})}.
    \label{eq:def_beta-tilde}
\end{align}
From the above observations, $\ket{\tilde{\beta}}$ is expected to be a good approximation for $\ket{\beta}$. In particular, expectation values in $\ket{\tilde{\beta}}$ of local observables whose supports are far from the boundary between subsystems $\Lambda_L$ and $\Lambda_R$ are expected to coincides with those in the {\BBB} state $\ket{\beta}$ in the thermodynamic limit.
Thus, to derive Eq.~\eqref{eq:formula_expectation-value} in the main text, for any local observable $\hat{O}$ with support around $r_1=L_1/4$, which is $O(L_1)$ distance from the boundary of $\Lambda_L$,
\begin{figure}[H]
    \centering
    \includegraphics{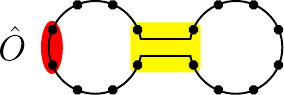}
\end{figure}
\noindent
we assume that $\ket{\beta}$ and $\ket{\tilde{\beta}}$ give the same expectation values in the thermodynamic limit:
\begin{align}
    \lim_{N\to\infty} \frac{\braket{\beta|\hat{O}|\beta}}{\braket{\beta|\beta}}
    = \lim_{N\to\infty} \frac{\braket{\tilde{\beta}|\hat{O}|\tilde{\beta}}}{\braket{\tilde{\beta}|\tilde{\beta}}}.
    \label{eq:assumption_expectation-value}
\end{align}
Additionally, to derive Eq.~\eqref{eq:formula_free-energy}, we assume that the logarithm of norms of $\ket{\beta}$ and $\ket{\tilde{\beta}}$ coincide in leading order of the thermodynamic limit:
\begin{align}
    \lim_{N\to\infty} \frac{1}{N} \log \braket{\beta|\beta}
    = \lim_{N\to\infty} \frac{1}{N} \log \braket{\tilde{\beta}|\tilde{\beta}}.
    \label{eq:assumption_free-energy}
\end{align}
The validity of these assumptions is tested numerically in the main text by verifying formulas~\eqref{eq:formula_free-energy} and \eqref{eq:formula_free-energy} derived from these assumptions.

\subsection{Properties of $\ket{\tilde{\beta}}$}
Let us examine the properties of $\ket{\tilde{\beta}}$ in detail.
We write the spin basis of the system on $\Lambda_{N/2}$, i.e., the tensor products of $N/2$ eigenvectors of $\hat{\sigma}^z$, as ${\ket{\vec{\sigma}}}$. Obviously, ${\ket{\vec{\sigma}}}$ form the basis for the Hilbert spaces associated with the subsystems on $\Lambda_L$ and $\Lambda_R$. Since the system on $\Lambda$ is a composite system consisting of two systems on $\Lambda_L$ and $\Lambda_R$, the EAP state $\ket{\mathrm{EAP}(\hat{u})}$ can be expanded using the spin basis as
\begin{align}
    \ket{\mathrm{EAP}(\hat{u})}
    = \hat{I}_L \otimes \left(\hat{u}^{\otimes N/2}\right) \bigotimes_{r \in \Lambda_L} \left[
        \ket{0}_{r}\ket{0}_{p(r)}+\ket{1}_{r}\ket{1}_{p(r)}    
    \right]
    = \sum_{\vec{\sigma}} \ket{\vec{\sigma}}_L \otimes \left(\hat{u}^{\otimes N/2}\right) \ket{\vec{\sigma}}_R.
\end{align}
Then, from Eqs.~\eqref{eq:Hamiltonian-decomposition} and \eqref{eq:def_beta-tilde}, we have
\begin{align}
    \ket{\tilde{\beta}}
    &= \sum_{\vec{\sigma}}
    e^{- \frac{1}{4} \beta \hat{H}_{N/2}} \ket{\vec{\sigma}}
    \otimes e^{- \frac{1}{4} \beta \hat{H}_{N/2}} \left(\hat{u}^{\otimes N/2}\right) \ket{\vec{\sigma}} \nonumber\\
    &= \sum_{\vec{\sigma}}
    e^{- \frac{1}{4} \beta \hat{H}_{N/2}} \ket{\vec{\sigma}}
    \otimes
    \left(\hat{u}^{\otimes N/2}\right) \left( \sum_{\vec{\sigma}'} {\ket{\vec{\sigma}'}\bra{\vec{\sigma}'}} \right) \left(\hat{u}^{\otimes N/2}\right)^\dagger e^{- \frac{1}{4} \beta \hat{H}_{N/2}} \left(\hat{u}^{\otimes N/2}\right) \ket{\vec{\sigma}} \nonumber\\
    &= \sum_{\vec{\sigma},\vec{\sigma}'}
    e^{- \frac{1}{4} \beta \hat{H}_{N/2}} \ket{\vec{\sigma}}
    \otimes
    \braket{\vec{\sigma}'|e^{- \frac{1}{4} \beta \left[ \left(\hat{u}^{\otimes N/2}\right)^\dagger \hat{H}_{N/2} \left(\hat{u}^{\otimes N/2}\right) \right]}|\vec{\sigma}} \left(\hat{u}^{\otimes N/2}\right) \ket{\vec{\sigma}'}.
    \label{eq:beta-tilde_purification_derivation}
\end{align}
Since $\hat{H}_{N/2}$ also has symmetry with respect to the antiunitary transformation $\hat{\Theta} = \hat{u}^{\otimes N/2} \hat{K}$, it holds that
\begin{align}
    {\braket{\vec{\sigma}|e^{- \frac{1}{4} \beta \left[ \left(\hat{u}^{\otimes N/2}\right)^\dagger \hat{H}_{N/2} \left(\hat{u}^{\otimes N/2}\right) \right]}|\vec{\sigma}'}}^*
    = \braket{\vec{\sigma}|e^{- \frac{1}{4} \beta \hat{H}_{N/2}}|\vec{\sigma}'}
\end{align}
for any $\ket{\vec{\sigma}}$ and $\ket{\vec{\sigma}'}$.
Therefore, substituting this into Eq.~\eqref{eq:beta-tilde_purification_derivation}, we get
\begin{align}
    \ket{\tilde{\beta}}
    &= \sum_{\vec{\sigma},\vec{\sigma}'}
    e^{- \frac{1}{4} \beta \hat{H}_{N/2}} \ket{\vec{\sigma}}\bra{\vec{\sigma}} e^{- \frac{1}{4} \beta \hat{H}_{N/2}} \ket{\vec{\sigma}'}
    \otimes \left(\hat{u}^{\otimes N/2}\right) \ket{\vec{\sigma}'}
    = \sum_{\vec{\sigma}'}
    e^{-\frac{1}{2}\beta\hat{H}_{N/2}} \ket{\vec{\sigma}'}
    \otimes \left(\hat{u}^{\otimes N/2}\right) \ket{\vec{\sigma}'}.
    \label{eq:beta-tilde_purification}
\end{align}

\subsection{Derivation of Eq.~\eqref{eq:formula_expectation-value}}
Now, let us derive the formula~\eqref{eq:formula_expectation-value} that gives thermal expectation values of local observables.

Since both $\ket{\beta}$ and $\hat{\rho}_N^\mathrm{can}(\beta)$ are translation invariant, without loss of generality, we can assume that $\hat{O}$ has support on $\Lambda_L$, particularly around $r_1=L_1/4$. Thus, we assume Eq.~\eqref{eq:assumption_expectation-value}.

On the other hand, from Eq.\eqref{eq:beta-tilde_purification}, for any observable $\hat{O}$ on $\Lambda_L$, we have
\begin{align}
    \braket{\tilde{\beta}|\hat{O}|\tilde{\beta}}
    &= \mathrm{Tr} \left[ e^{-\beta\hat{H}_{N/2}} \hat{O} \right],\\
    \braket{\tilde{\beta}|\tilde{\beta}}
    &= \mathrm{Tr} \left[ e^{-\beta\hat{H}_{N/2}} \right]. \label{eq:beta-tilde_norm}
\end{align}
Therefore, we have
\begin{align}
    \frac{\braket{\tilde{\beta}|\hat{O}|\tilde{\beta}}}{\braket{\tilde{\beta}|\tilde{\beta}}}
    &= \mathrm{Tr} \left[ \hat{\rho}_{N/2}^\mathrm{can}(\beta) \hat{O} \right],
\end{align}
where $\hat{\rho}_{N/2}^\mathrm{can}(\beta) \propto e^{-\beta\hat{H}_{N/2}}$ is the Gibbs state for $\hat{H}_{N/2}$.
Combining this with assumption~\eqref{eq:assumption_expectation-value}, in the thermodynamic limit, we obtain
\begin{align}
    \lim_{N\to\infty} \frac{\braket{\beta|\hat{O}|\beta}}{\braket{\beta|\beta}}
    &= \lim_{N\to\infty} \mathrm{Tr} \left[ \hat{\rho}_N^\mathrm{can}(\beta) \hat{O} \right].
\end{align}
Therefore, we have Eq.~\eqref{eq:formula_expectation-value}.

\subsection{Derivation of Eq.~\eqref{eq:formula_free-energy}}
Finally, let us derive the formula~\eqref{eq:formula_free-energy} for the free energy density. From Eq.~\eqref{eq:beta-tilde_norm}, we immediately get
\begin{align}
    \frac{2}{N} \log \braket{\tilde{\beta}|\tilde{\beta}}
    = \frac{1}{N/2} \log Z_{N/2},
\end{align}
where $Z_{N/2} = \mathrm{Tr} \left[ e^{-\beta\hat{H}_{N/2}} \right]$ is the partition function for the system with Hamiltonian $\hat{H}_{N/2}$. Hence the thermodynamic limit yields
\begin{align}
    \lim_{N\to\infty} \frac{2}{N} \log \braket{\tilde{\beta}|\tilde{\beta}}
    = - \beta f(\beta)
\end{align}
Therefore, combining this with assumption~\eqref{eq:assumption_free-energy}, we obtain
\begin{align}
    \lim_{N\to\infty} \frac{2}{N} \log \braket{\beta|\beta}
    = - \beta f(\beta).
\end{align}
Therefore, we have Eq.~\eqref{eq:formula_free-energy}.

It is worth mentioning that this is consistent with the results of field theoretical analysis for $(1+1)$-dimensional CFTs, where EAP states correspond to crosscap states.
Consider a (perturbed) CFT defined on a spatial circle of circumference $L$.
According to the results of Refs.~\cite{Tu2017,Zhang2023,Tan2024}, the cylinder partition function with crosscap boundary states,
\begin{align}
    Z^\mathcal{C}(L,\beta/2)
    = \braket{\mathrm{EAP}(\hat{u})|e^{-\frac{1}{2}\beta\hat{H}}|\mathrm{EAP}(\hat{u})}
    (= \braket{\beta|\beta}),
\end{align}
coincides with the partition function $Z^\mathcal{K}(L/2,\beta)$ on the Klein bottle, a rectangular $[0,L/2] \times [0,\beta]$ with sides identified by the relations $(x,0) \sim (x,\beta)$ for $0 \leq x \leq L/2$ and $(0,\tau) \sim (L/2,\beta-\tau)$ for $0 \leq \tau \leq \beta$,
\begin{align}
    Z^\mathcal{C}(L,\beta/2) = Z^\mathcal{K}(L/2,\beta).
\end{align}
Furthermore, the ratio between $Z^\mathcal{K}(L/2,\beta)$ and the partition function $Z^\mathcal{T}(L/2,\beta)$ on the torus, which is the usual partition function $Z_{L/2}$ for a system of length $L/2$ at the inverse temperature $\beta$, approaches a constant independent of $L$ in the thermodynamic limit,
\begin{align}
    \frac{Z^\mathcal{K}(L/2,\beta)}{Z^\mathcal{T}(L/2,\beta)} = O(L^0).
\end{align}
Therefore, we have
\begin{align}
    \lim_{L\to\infty} \frac{2}{L} \log \braket{\beta|\beta}
    = \lim_{L\to\infty} \frac{1}{L/2} \log Z_{L/2}.
\end{align}
This is the field-theoretic counterpart of Eq.~\eqref{eq:formula_free-energy}.
Our claim in this subsection is that this holds in general, not just for $(1+1)$-dimensional CFT systems.

\newpage
\section{Dynamical {\BBB} state algorithm}
The {\BBB} state can also correctly describe dynamics starting from an equilibrium state. However, as time evolves, the {\BBB} state loses its simple entanglement structure, making it hard to represent efficiently as a tensor network state and challenging to compute efficiently over long timescales. To clarify this, let us consider the example of calculating a time-dependent correlation function.
The time-dependent correlation function of local observables $\hat{A}$ and $\hat{B}$ is defined as
\begin{align}
    C(t) = \lim_{N\to\infty} \mathrm{Tr} \left[ \hat{\rho}_N^\mathrm{can}(\beta) \hat{A}(t) \hat{B} \right],
\end{align}
where $\hat{A}(t) = e^{+i\hat{H}t} \hat{A} e^{-i\hat{H}t}$ is the Heisenberg operator for $\hat{A}$. Such time-dependent quantities can also be computed from the {\BBB} state in a manner analogous to the algorithm developed for the METTS method~\cite{Binder2015,Wang2024}.

In systems with short-range interactions, the Heisenberg operator for a local observable with $t$ independent of $N$ can be well approximated as a local observable~\cite{Yoneta2023_stationarity}. Consequently, in the same manner as for Eq.~\eqref{eq:formula_expectation-value}, we can show that for $t\in\mathbb{R}$ independent of $N$, the following holds:
\begin{align}
    C(t) = \lim_{N\to\infty} \frac{\braket{\beta| \hat{A}(t) \hat{B} |\beta}}{\braket{\beta|\beta}}.
\end{align}
To numerically calculate the right-hand side of the above equation, we proceed as follows. First, construct the {\BBB} state $\ket{\beta}$ using the algorithm described in the main text.
With this $\ket{\beta}$, set
\begin{align}
    \ket{\varphi_1(0)} = \frac{1}{\sqrt{\braket{\beta|\beta}}} \hat{B} \ket{\beta}, \qquad
    \ket{\varphi_2(0)} = \frac{1}{\sqrt{\braket{\beta|\beta}}} \ket{\beta}.
\end{align}
Next, evolve these states forward in time as
\begin{align}
    \ket{\varphi_1(t)} = e^{i\hat{H}t} \ket{\varphi_1(0)}, \qquad
    \ket{\varphi_2(t)} = e^{i\hat{H}t} \ket{\varphi_2(0)}.
\end{align}
Finally, compute the product $\braket{\varphi_2(t)|\hat{A}|\varphi_1(t)}$.
Then, since
\begin{align}
    \braket{\varphi_2(t)|\hat{A}|\varphi_1(t)}
    = \frac{\braket{\beta| \hat{A}(t) \hat{B} |\beta}}{\braket{\beta|\beta}},
\end{align}
the time-dependent correlation function $C(t)$ can be computed as
\begin{align}
    C(t) = \lim_{N\to\infty} \braket{\varphi_2(t)|\hat{A}|\varphi_1(t)}.
\end{align}

As explained in the main text, by rearranging the lattice, $\ket{\beta}$ can be transformed into a state with low entanglement. Furthermore, on the rearranged lattice, $\hat{H}$ retains short-range interactions. Therefore, for short timescales, both $\ket{\varphi_1(t)}$ and $\ket{\varphi_2(t)}$ exhibit low entanglement and can be efficiently represented as tensor network states. However, as time evolves, the {\BBB} state is expected to develop a more intricate entanglement structure, making it increasingly difficult to express as a tensor network state (see the example below).

\begin{figure*}[!h]
    \centering
    \subfloat[time-dependent correlation function $\braket{\hat{\sigma}_{1}^z(t)\hat{\sigma}_{1}^z}$]{\includegraphics[height=6cm]{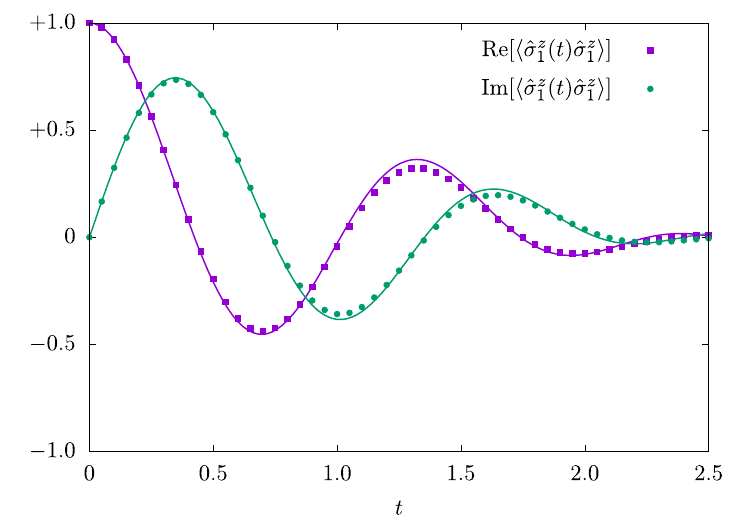}\label{fig:XY_time-dependent_correlation}} \hfill
    \subfloat[entanglement entropy $S_\mathrm{vN}$]{\includegraphics[height=6cm]{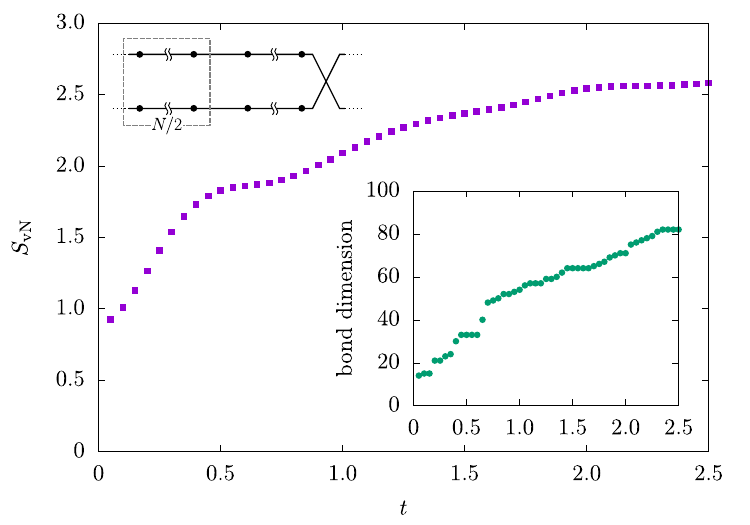}\label{fig:XY_time-dependent_EE}}
    \caption[]{%
        Dynamical properties of the {\BBB} state for the one-dimensional XY model with $J^{xx}=1, J^{yy}=1/2$ and $N=100$.
        \subref{fig:XY_time-dependent_correlation} Time-dependent correlation function $\braket{\hat{\sigma}_{1}^z(t)\hat{\sigma}_{1}^z}$ at the inverse temperature $\beta=1$. The solid line represents the exact solution in the thermodynamic limit.
        \subref{fig:XY_time-dependent_EE} Entanglement entropy of the time evolved {\BBB} state $\ket{\varphi_2(t)}$ at the inverse temperature $\beta=1$ for a center cut of the ladder. The inset shows the bond dimension of the MPS at the center cut.
        Real- and Imaginary-time evolution are carried out using a second-order Trotter decomposition with a time step of $0.05$. We set a truncation threshold of $10^{-6}$.
    }
    \label{fig:XY_time-dependent}
\end{figure*}

As a demonstration, we apply the algorithm explained above to the the one-dimensional XY model (defined by Eq.~\eqref{eq:XY}) with $N=100$
and compute the time-dependent correlation function of the Pauli matrix $\hat{\sigma}_{1}^z$,
\begin{align}
    C(t) = \lim_{N\to\infty} \mathrm{Tr} \left[ \hat{\rho}_N^\mathrm{can}(\beta) \hat{\sigma}_{1}^z(t) \hat{\sigma}_{1}^z \right].
\end{align}
In Fig.~\ref{fig:XY_time-dependent_correlation}, we plot the time-dependent correlation function $C(t)$ computed from the {\BBB} state. The numerical results show close agreement with the exact solution~\cite{Niemeijer1967}, with only slight deviations. These small discrepancies arise from using a relatively large truncation threshold of $10^{-6}$ to reduce computational costs.
However, as illustrated in Fig.~\ref{fig:XY_time-dependent_EE}, the bipartite entanglement of the time evolved {\BBB} state $\ket{\varphi_2(t)}$ on the rearranged lattice (ladder) grows with time, resulting in an increase in the bond dimension of the MPS. This severely degrades computational efficiency.
Thus, while the {\BBB} state correctly provides dynamical properties, it is practically challenging to perform numerical simulations over long timescales.

\end{document}